\documentclass[aps,prl,showpacs, floatfix,twocolumn,amsmath,amssymb, superscriptaddress]{revtex4}

\usepackage{graphicx}
\usepackage{hyperref}
\usepackage[usenames]{color}

\begin{document}

\title{Influence of topological excitations on Shapiro steps and microwave dynamical conductance in bilayer exciton condensates}

\author{Timo Hyart}
\affiliation{Institut f\"ur Theoretische Physik, Universit\"at Leipzig, D-04103, Leipzig, Germany}
\author{Bernd Rosenow}
\affiliation{Institut f\"ur Theoretische Physik, Universit\"at Leipzig, D-04103, Leipzig, Germany}



\begin{abstract}
The quantum Hall state at total filling factor $\nu_T=1$ in bilayer systems realizes an exciton condensate and exhibits a zero-bias tunneling anomaly, similar to the Josephson effect in the presence of fluctuations. In contrast to conventional Josephson junctions,  no Fraunhofer diffraction pattern has  been observed, due to  disorder induced topological defects, so-called merons. We consider interlayer tunneling  in the presence of microwave radiation, 
and  find  Shapiro steps in the tunneling current-voltage characteristic despite the presence of merons. Moreover, the Josephson oscillations can also be observed as resonant features in the microwave dynamical conductance.
\end{abstract}

\maketitle

Quantum Hall (QH) systems are an ideal platform to study the rich diversity of exotic
effects induced by Coulomb interactions  \cite{Perspectives, Girvin-summer-school}. One state of particular
interest is an exciton condensate  occurring in QH bilayers at the total
filling factor $\nu_T = 1/2 + 1/2 = 1$  \cite{Perspectives, Girvin-summer-school,  Eisenstein-MacDonald, fertig89, Moon95, WenandEzawa, Stern01, Hyart, Fertig03, Eastham09, Eastham10, Su10, Spielman00,Spielman01,quantized-Hall-Drag1,
Spielman-diss,Kellog04,Wiersma04-06,Finck08, TiemannPRB09,Yoon, Corbino-Eisenstein, Xuting}.  
 Recently, there has been growing interest in similar exciton condensate states in  bilayer graphene \cite{Min08, Seradjeh-graphene} and topological insulator thin films \cite{Seradjeh-TIfilms}, and in this Letter the QH bilayer state serves as a prototype system for exciton condensates with interesting topological excitations. Although the excitons are charge neutral, the  bilayer nature of these states allows remarkable electronic properties such as counterflow
superconductivity and a tunneling supercurrent similar to the
Josephson current between two superconductors \cite{Perspectives, Girvin-summer-school,  Eisenstein-MacDonald, fertig89, Moon95, WenandEzawa, Stern01, Hyart, Fertig03, Eastham09, Eastham10, Su10}. 
A
spectacular enhancement of the tunneling conductance at small
interlayer bias voltage, a quantized Hall drag resistance, and almost dissipationless counterflow
currents have been observed experimentally \cite{Spielman00,Spielman01,quantized-Hall-Drag1,
Spielman-diss,Kellog04,Wiersma04-06,Finck08, TiemannPRB09,Yoon, Corbino-Eisenstein, Xuting}.

The macroscopic phase coherence in Josephson junctions (JJs) was  originally confirmed by measuring the dependence of the tunneling current on  magnetic field \cite{Rowell} and microwave radiation \cite{Shapiro}. In the first type of experiment, oscillations of the critical current were observed  as a function of the magnetic flux $\Phi$ applied across the junction.
In particular, for an integer number of superconducting flux quanta $h/2e$ in the junction, the tunneling current vanishes, in 
analogy to the Fraunhofer diffraction pattern in the case of light passing through a narrow rectangular slit. In the second type of experiment, the application of the microwave radiation allows the observation of the ac Josephson effect. Namely, in the presence of a dc voltage and microwave radiation the phase difference between the two superconductors oscillates at two characteristic frequencies, the
Josephson frequency determined by the dc voltage $2eV_{dc}/\hbar$ and the microwave frequency $\omega$, and the interplay of these  oscillations gives rise to steps in the current-voltage (I-V) characteristic whenever the dc voltage satisfies $2eV_{dc}=n \hbar\omega$ ($n=0,1,2,...$).

In QH  bilayer exciton condensates, a weak tunnel coupling between the layers gives rise to a  Josephson-like tunneling effects  \cite{WenandEzawa}, with the Cooper pair charge $2e$ replaced by the usual electron charge $e$. However, the QH exciton condensate supports exotic topological excitations, so-called merons. They are vortices of the order parameter field and carry a charge $\pm e/2$  localized in one of the layers  \cite{Moon95, Perspectives, Girvin-summer-school}.   Because they are charged, merons can be nucleated by a disorder potential, and thus give rise to a correlation length 
$\xi$ for the condensate phase, determined by the average distance between merons.
Small domains of characteristic size $\xi^2$ act as individual JJs, well-coupled to each other by counterflow currents.  As the size of these domains varies randomly, averaging over it 
"washes away" the Fraunhofer diffraction pattern \cite{Hyart, Spielman-diss} and gives rise to a smooth decay
of the tunneling current with the in-plane magnetic field.  Based on this result one might expect that merons also destroy the Shapiro steps in QH bilayers. We address this question by studying the dynamics of a QH exciton condensate in the presence of  microwave radiation. 
We find that  Shapiro-like steps in the I-V characteristic are indeed present at  dc voltages 
\begin{equation}
eV_{dc}=n\hbar \omega, \hspace{0.8 cm} (n=0, 1, 2,...) 
\end{equation}
despite the presence of the merons. Moreover, we show that  Josephson oscillations also result in resonant features in the microwave dynamical conductance, and lead to regions of absolute negative conductance (ANC).

We concentrate on  tunneling deep inside the coherent phase at $\nu_T=1$, 
and assume that the real spin is fully polarized. Then, the low energy theory can be formulated in terms of a pseudospin, which describes the \textit{which layer} quantum degree of freedom and can also be considered as the exciton condensate order parameter \cite{Moon95,Perspectives, Girvin-summer-school}. The low-energy excitations of the system are the pseudospin waves and the merons. In the absence of tunneling,  pseudospin waves are described by the Hamiltonian density \cite{Moon95,Perspectives,Girvin-summer-school, Stern01}
\begin{equation}
 \mathcal{H}=\frac{1}{2} \rho_s (\nabla \varphi)^2+\frac{(e n_0 m_z/2)^2}{2 \Gamma}, \label{startingpoint}
\end{equation}
where $\vec{m}=(\cos \varphi, \sin \varphi, m_z)$ is the pseudospin vector, $\rho_s$  the pseudospin stiffness, $\Gamma$  the capacitance per area and $n_0$ is the average density. The momentum conjugate to
$\varphi$ is $\pi=\hbar n_0 m_z/2$. The first term in Eq.~(\ref{startingpoint}) arises from the loss of Coulomb exchange energy if the pseudospin direction varies in space, and the second term measures the capacitive energy \cite{Moon95, Stern01}. The dispersion relation of pseudo spin waves described by the 
 Hamiltonian (\ref{startingpoint}) is
$\omega_{\vec{k}}=u k$, with
$u=\sqrt{\rho_s e^2/\Gamma \hbar^2}$ the spin wave velocity.

We consider a homogeneous time-dependent interlayer voltage $V(t)$. The tunneling energy and current operators are $H_T=T_++T_-$ and $\hat{I}=i e (T_+-T_-)/\hbar$, where
$$
 T_{\pm}(t)=- \lambda e^{\pm i e\int_{t_0}^t V(t')dt'/\hbar} \int d^2 r \ e^{\pm iQ_B x} e^{\pm i \varphi} e^{\pm i \varphi_m}.
$$
Here $\lambda=\Delta_{SAS}/(8 \pi l_B^2)$, $\Delta_{SAS}$ is the tunnel coupling between the two layers, 
$l_B=\sqrt{\hbar/e B}$ is the magnetic length, 
$Q_B=eB_{||}d/\hbar$, $B_{||}$ is the in-plane magnetic field and $d$ is the separation between the layers.  Merons are included phenomenologically with the help of the vortex field $\varphi_m$ \cite{Stern01, Hyart, Eastham10}. Both counterflow and tunneling experiments suggest that fluctuations of $\varphi_m$  are important. We characterize the vortex field fluctuations by a correlation length $\xi$ and a correlation time $\tau_\varphi$,  such that
 $\langle e^{\pm i \varphi_m(\vec{r}_1, t_1)} e^{\mp i \varphi_m(\vec{r}_2, t_2)} \rangle  =e^{-|\vec{r}_1-\vec{r}_2|/\xi}e^{-|t_1-t_2|/\tau_\varphi}$ and
$\langle e^{\pm i \varphi_m(\vec{r}_1, t_1)} e^{\pm i \varphi_m(\vec{r}_2, t_2)} \rangle =0$.
These assumptions  give rise to a quantitative agreement between theory and experiments, 
see Ref.~\cite{Hyart}. 

In the presence of vortex field fluctuations and for a small tunneling amplitude,  tunneling acts as a bottleneck in charge transport, and the time-dependent tunneling current can be calculated perturbatively using linear response theory
 \begin{equation}
I(t)=\frac{i}{\hbar} \int_{t_0}^t  dt' <[H_T(t'), \hat{I}(t)]>. 
\end{equation}
The solution for arbitrary time-dependent voltage
\begin{equation}
V(t)=V_{dc} +\sum_{k=1}^{N} V_{\omega_{k}}\cos(\omega_{k}t +
\alpha_{k}) \label{E-field-uudestaan}
\end{equation}
can be computed  by  exploiting the relation
%
\begin{equation}
\langle e^{i \varphi(\vec{r},t)} e^{-i \varphi(0,0)} \rangle = \ e^{\frac{1}{2} [\varphi(\vec{r},t), \varphi(0,0)]} \ e^{- {1 \over 2} \langle (\varphi(\vec{r},t) - \varphi(0,0))^2 \rangle}, 
\end{equation}
%
valid for any quadratic Hamiltonian. In this way,
we obtain
\begin{eqnarray}
I(t)&=&\sum_{n_1,...,n_N}
\sum_{m_1,...,m_N} \bigg[ \prod_{k=1}^N J_{n_k} (
\beta_k) J_{n_k+m_k}( \beta_k) \bigg] \nonumber\\ && \hspace{-0.5cm} \times \bigg\{
I_{S} \big(eV_{dc}+\sum_{k=1}^N n_k \hbar \omega_k \big)
\cos{\big[\sum_{k=1}^N m_k(\omega_k t+\alpha_k)\big]}\nonumber
\\&& \hspace{-0.5cm}+K_S \big(eV_{dc}+\sum_{k=1}^{N} n_k \hbar \omega_k \big)
\sin{\big[\sum_{k=1}^N m_k(\omega_k t+\alpha_k)\big]} \bigg\}, \nonumber \\
\label{Tucker-formulas}
\end{eqnarray}
where $J_n(x)$ are Bessel functions, the summations are from $-\infty$ to $\infty$ and $\beta_k=eV_{\omega_k}/\hbar \omega_k$. Here, $I_S(e V_{dc})= \textrm{Im}[F_S(eV_{dc}) ]$ is the static I-V characteristic in the absence of microwave field \cite{Stern01, Hyart}, and   $K_S(eV_{dc})=\textrm{Re}[F_S(eV_{dc})]$  is the real part of the  complex function 
\begin{eqnarray}
F_S(eV)&=& \int d q   \bigg[\frac{I_0}{q-eV/eV_0-i \alpha} +\frac{I_0}{eV/eV_0+q+i \alpha} \bigg] \nonumber \\ &&\times \int dR \ R \ e^{-R} J_0(q R) J_0(Q_B \xi R).
\label{FS}
\end{eqnarray}
Here $q=k\xi$ and $R=r/\xi$ are dimensionless momentum and radial coordinate,
$
I_0= (e \xi^2 L^2 \Delta_{SAS}^2 e^{-D_0/2})/( 64 \pi^2 \hbar \rho_s l_B^4) 
$
and $V_0 = \hbar u /e\xi$ determine the characteristic current and voltage scales,
$\alpha(T) =\xi/u \tau_\varphi(T)$ is the temperature dependent decoherence rate,
$D_0 = \hbar u k_0/(2\pi \rho_s)$, $k_0 = \kappa \sqrt{2}/l_B$  ($\kappa \approx 1$) is an ultraviolet cut-off momentum and $L^2$ is the area of the sample. Because $I_S$ and $K_S$  are related to each other by the Kramers-Kronig relations, the response to arbitrary ac field is completely determined by the static I-V characteristic. 

The I-V characteristic obtained from Eq.~(\ref{FS}) can be used to describe a large number of experimental observations \cite{Hyart}. In particular, by fitting  $\xi  \sim 100$ nm and  $u \sim 14$ km/s, the smooth suppression of the conductance peak without the Fraunhofer oscillations  and an appearance of the resonant enhancement of the tunneling current at $e V_{dc}=\hbar uQ_B$ in the presence of the in-plane magnetic field can be quantitatively described. Moreover, the temperature dependent decoherence rate $\alpha(T)$ can be used to explain the height and width of the zero-bias conductance peak.
 The parameters $V_0$ and $I_0$ depend only weakly on temperature, and have typical values  $V_0 \sim 100$ $\mu$V and $I_0 \sim 5$ pA - $1$ nA  \cite{Hyart}. The current scale $I_0$ can vary considerably, because the tunnel coupling $\Delta_{SAS}$ can be experimentally controlled over a wide range of values $10$ - $100$ $\mu$K.  For temperatures comparable to the thermal activation gap, the temperature dependence of $\alpha$ arises mainly due to thermally activated hopping of the merons. On the other hand, at smaller temperatures $\alpha$  shows a power-law like temperature dependence, which could originate from low-energy excitation close the the meron cores \cite{Fertig03}. At lowest experimental temperatures we obtain $\alpha \approx 0.01$ \cite{Hyart}, which means that $\tau_\varphi \sim 1$ ns. Therefore we  expect interesting features in the microwave response at GHz frequencies. In the following we measure the frequency in scaled units as $\hbar \omega/eV_0$. A frequency $\omega/2\pi=1$ GHz  corresponds to $\hbar \omega/eV_0 \approx 0.04$.

\begin{figure}
\includegraphics[scale=0.37]{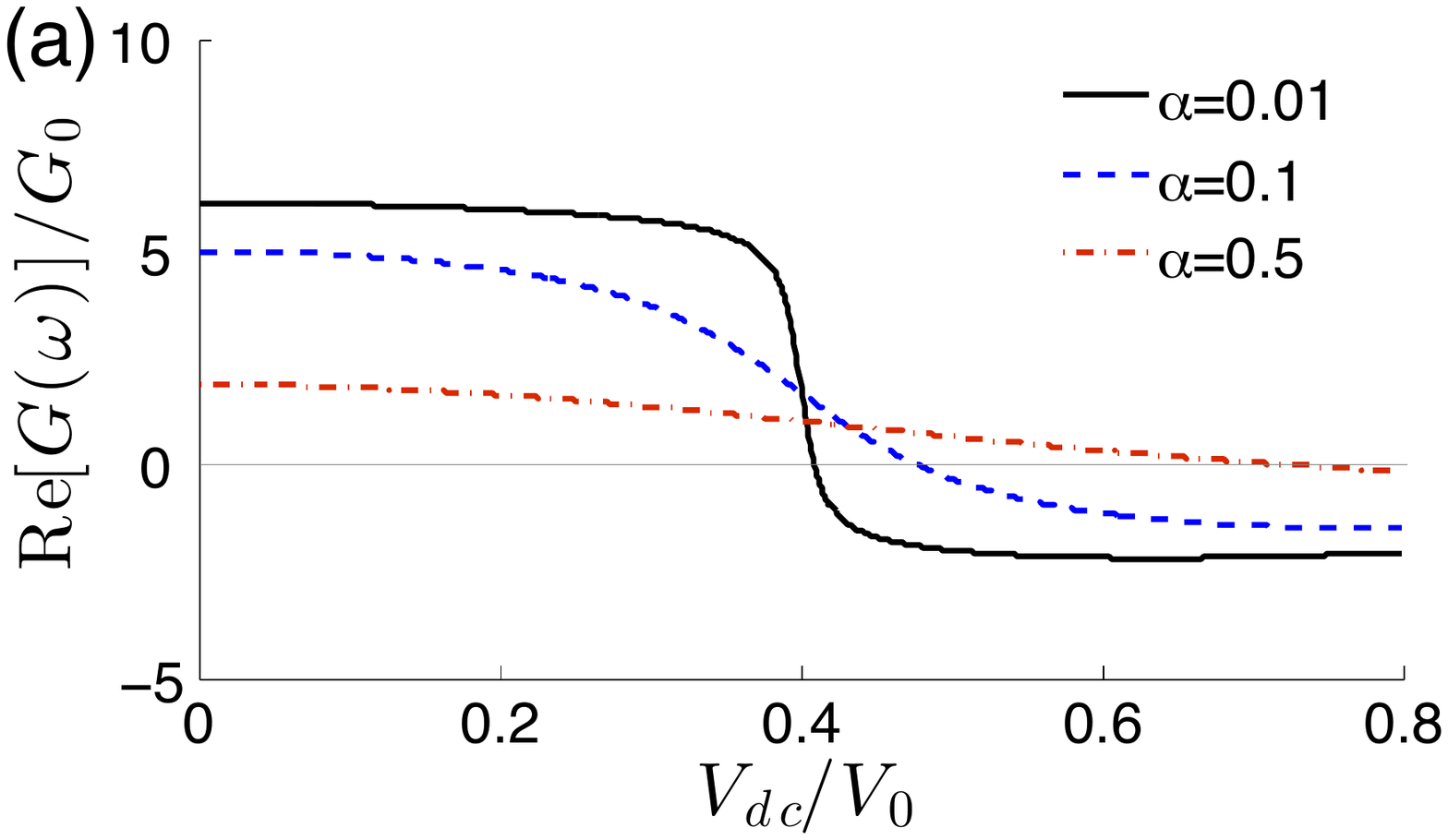} \includegraphics[scale=0.37]{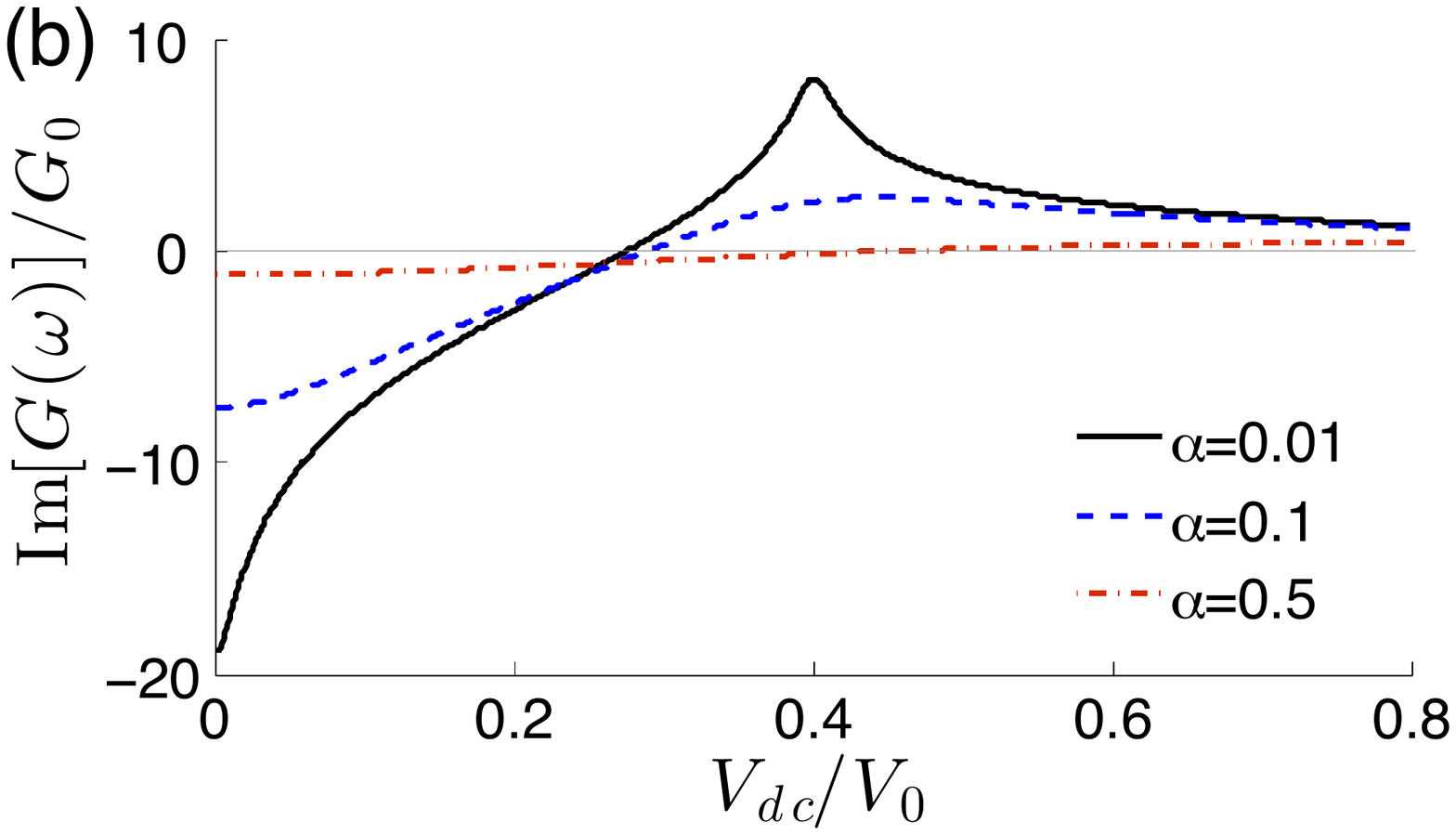} 
\caption{(color online) (a) Real and (b) imaginary parts of the small signal dynamical conductance for $\hbar \omega/eV_0=0.4$ and different dissipation strengths $\alpha$. The conductance is given in units $G_0=I_0/V_0$. The $I_1^c=0$ and $I_1^s=0$ are marked with a thin gray line.} \label{Dyncond}
\end{figure}

The relations (\ref{Tucker-formulas}) are generalization of the so-called Tucker relations to arbitrary polychromatic field, and have interesting consequences in various tunneling systems  \cite{Tucker, Wacker, Platero, Photon-assisted-transport, Falci}.  For slowly varying voltage,  $\omega_i \tau_\varphi \ll1$, the tunneling current follows the instantaneous value of the time-dependent voltage according to the static I-V characretistic $I(t)=I_S(eV(t))$,
but for $\omega_i \tau_\varphi \gtrsim 1$ significant deviations from this relation can occur. Here,
we concentrate on the situation where the microwave field is monochromatic $V(t)=V_{dc} +V_{\omega}\cos\omega t$. In this case, the dc component $I_{dc}$ and the ac components $I_m^{c}$, $I_m^s$
of the time-dependent current can be written as
\begin{eqnarray}
I(t)&=&I_{dc}+\sum_{m=1}^\infty [I_m^{c} \cos(m \omega t)+I_m^{s} \sin(m\omega t)],  \nonumber\\
I_{dc} &=&\sum_{n} J_n^2 (\beta) I_S(eV_{dc}+n\hbar \omega),  \nonumber  \\
I_m^{c}  &=& \sum_{n}
J_n(\beta)[J_{n+m}(\beta) +J_{n-m}(\beta)]  I_S (eV_{dc}+n \hbar
\omega), \nonumber
 \\
I_m^{s} &=& \sum_{n}
J_n(\beta)[J_{n+m}(\beta)-J_{n-m}(\beta)] K_S(eV_{dc}+n \hbar
\omega), \nonumber \\
\label{I-monochromatic}
\end{eqnarray}
where the summations are from $-\infty$ to $\infty$, and $\beta=eV_\omega/\hbar\omega$.
Defining the  dynamical conductance via $I_\omega = G(\omega) V_\omega$, we can identify
${\rm Re} [G(\omega)] = I^c_1 / V_\omega$, and  ${\rm Im} [G(\omega)] = -I^s_1 / V_\omega$.
Because all the microscopic parameters are known, we can use Eqs.~(\ref{I-monochromatic}) to calculate the microwave response of the QH exciton condensates. 
\begin{figure}
\includegraphics[scale=0.37]{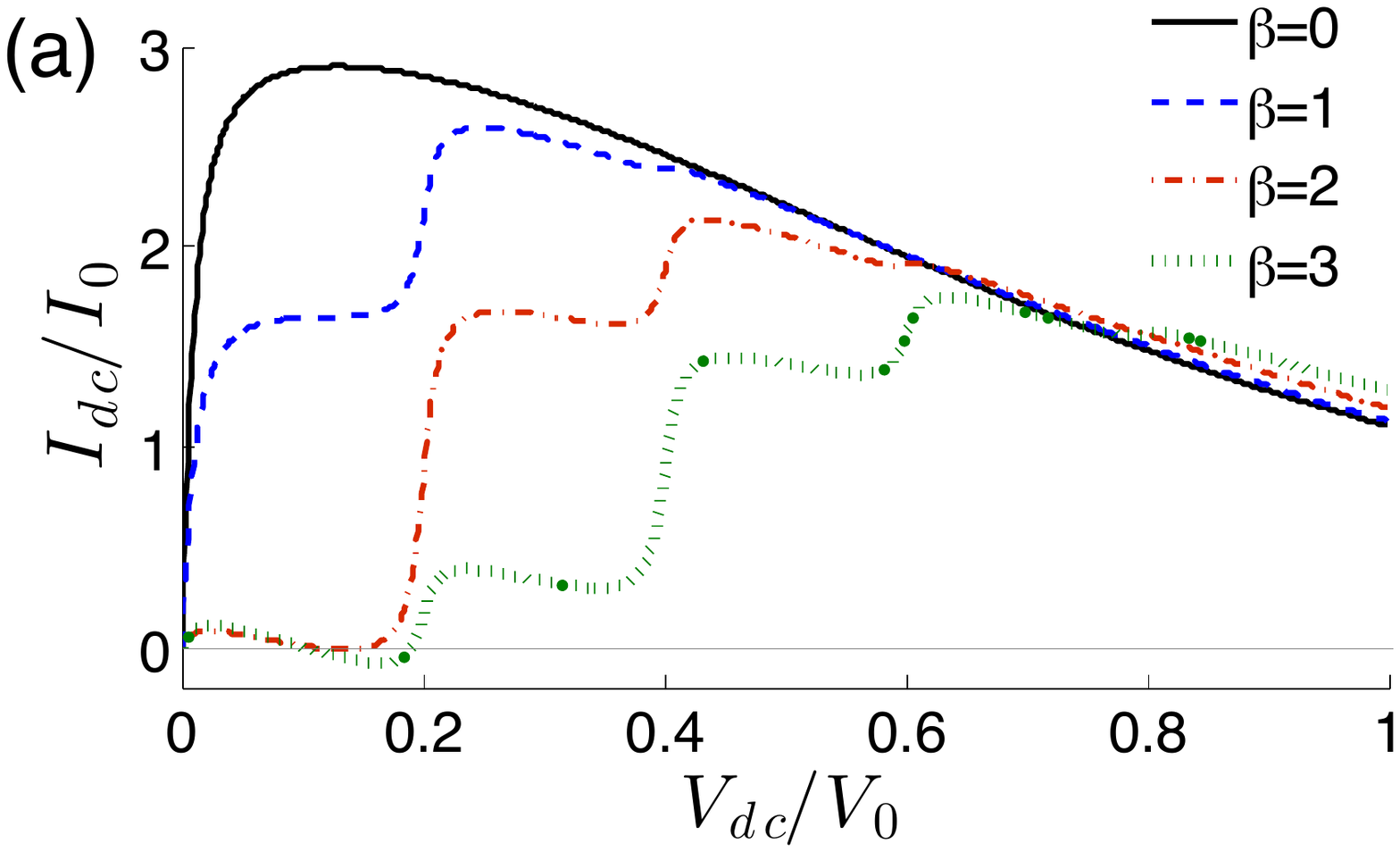} \includegraphics[scale=0.37]{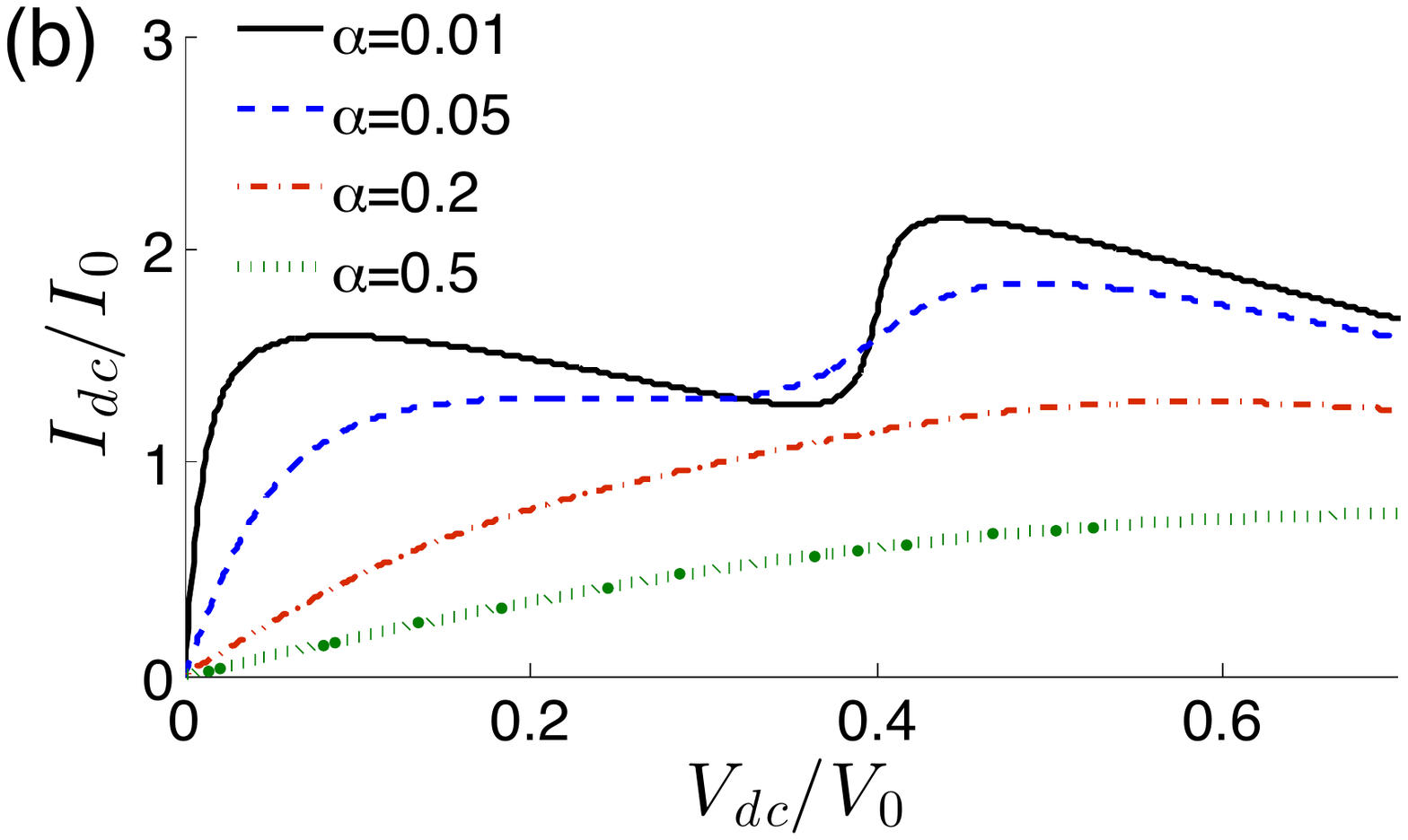}
\includegraphics[scale=0.37]{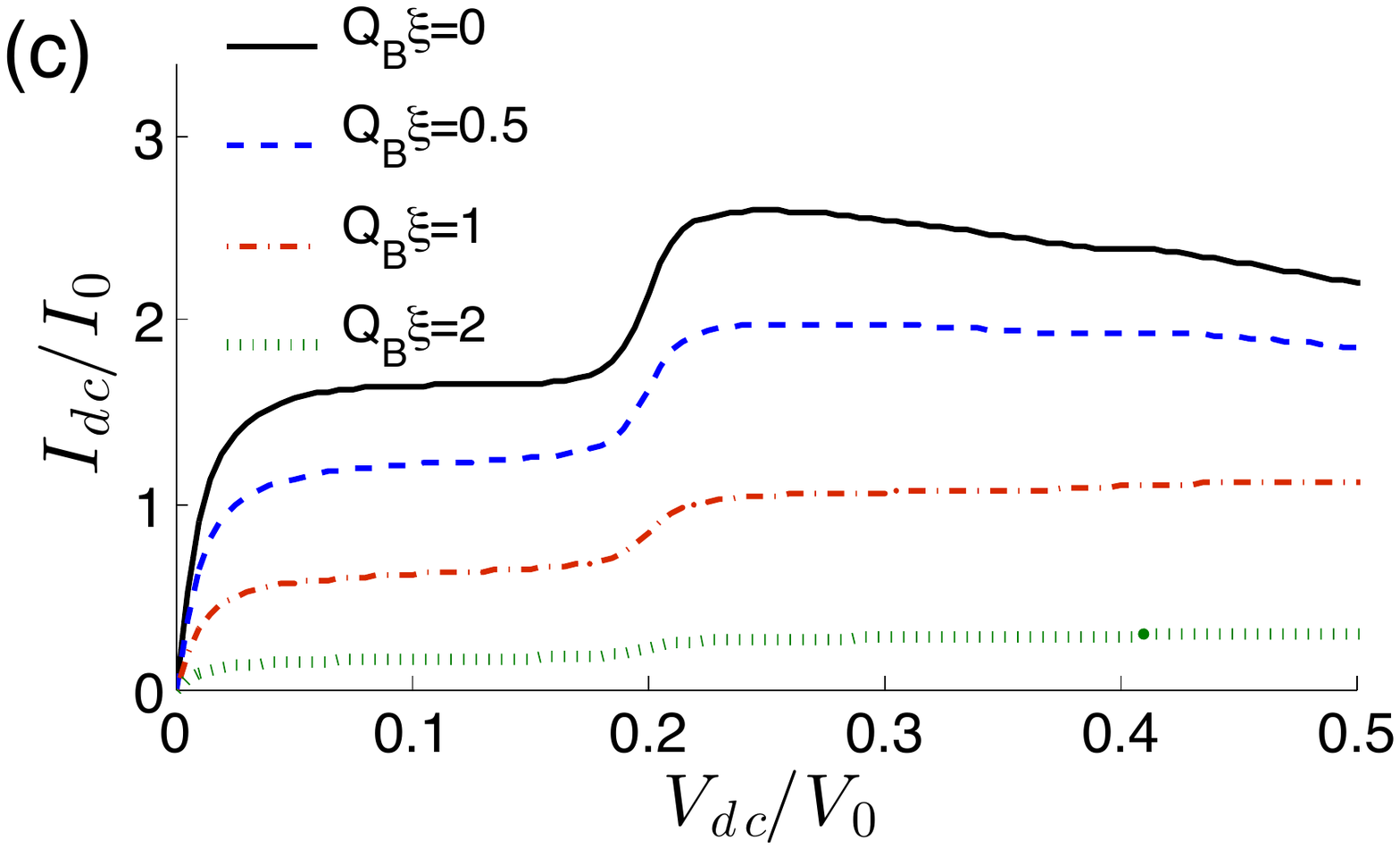} \includegraphics[scale=0.37]{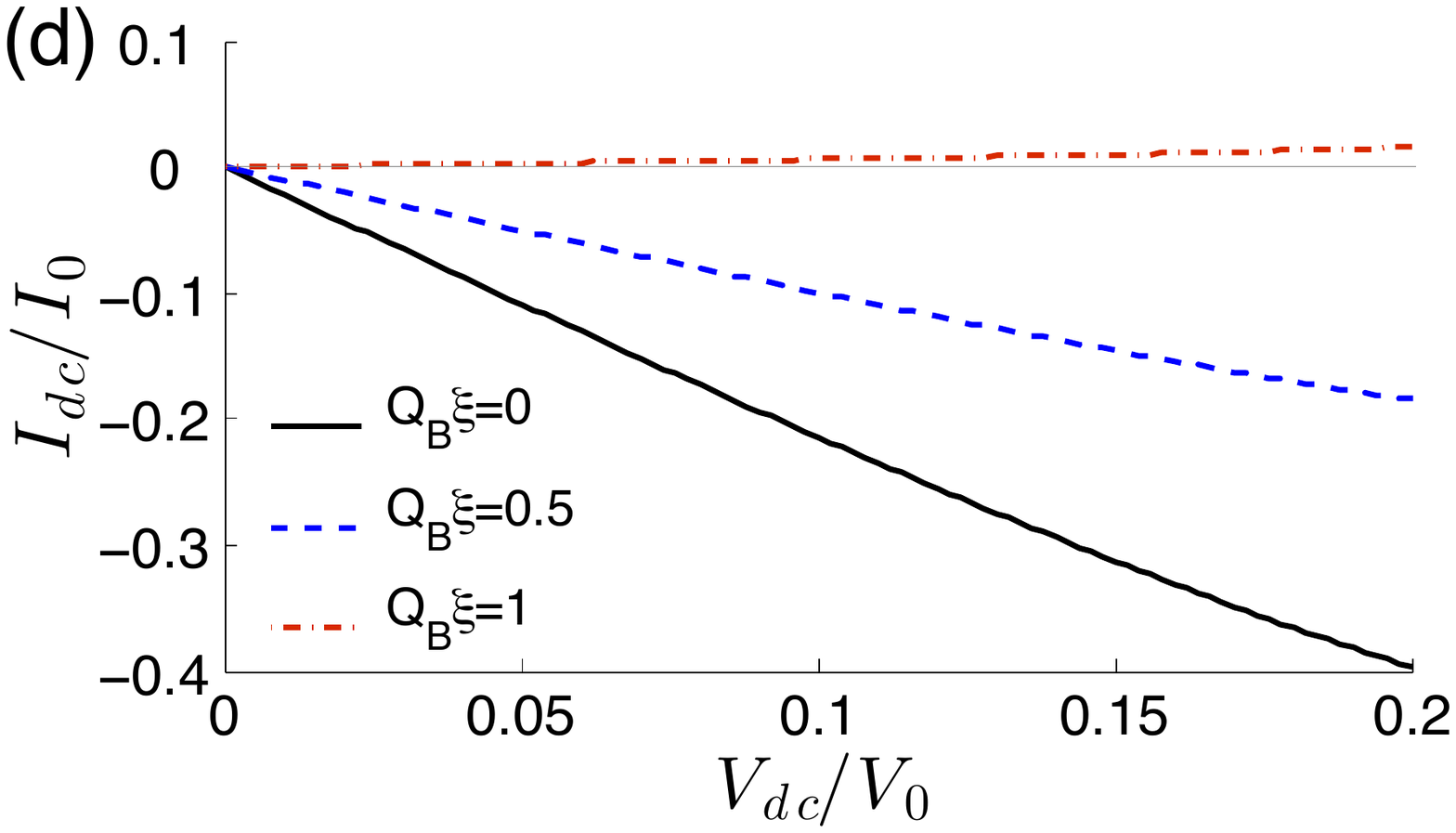}
\caption{(a) (color online) The I-V characteristics demonstrating  Shapiro steps at voltages $e V_{dc}=n \hbar \omega$ ($n=0, 1, 2,...$)  for $\alpha=0.01$, $\hbar \omega/eV_0=0.2$, $Q_B \xi=0$ and various $\beta=e V_\omega/\hbar \omega$.   (b) Dependence of the I-V characteristics on the dissipation strength $\alpha$ for $\hbar \omega/eV_0=0.4$, $\beta=1$ and $Q_B \xi=0$. (c) Suppression of  Shapiro steps in the presence of the in-plane magnetic field for $\alpha=0.01$, $\hbar \omega/eV_0=0.2$ and $\beta=1$. (d) Demonstration of absolute negative conductance for $\alpha=0.01$, $\hbar \omega/eV_0=0.3$ and $\beta=2.4$. By increasing the in-plane magnetic field, the direction of the current can be reversed. $I_{dc}=0$ is marked with a thin gray line in (a) and (d).}
\label{FigShapirosteps}
\end{figure}
Our results for the I-V characteristics and dynamical conductance for different values of the dissipation strength, in-plane magnetic field, and the amplitude and frequency of the microwave field are shown in Figs.~\ref{FigShapirosteps} and \ref{Dyncond}. Clear step-like structures appear in the I-V characteristic if $\hbar \omega/eV_0 \gg \alpha$. Therefore for sufficiently low dissipation strength $\alpha=0.01$, the Shapiro steps can be observed at GHz frequencies. As can be seen from Eqs.~(\ref{I-monochromatic}) and Figs.~\ref{FigShapirosteps} (a), (b) and (c), the Shapiro steps appear at voltages $e V_{dc}=n \hbar \omega$ ($n=0, 1, 2,...$) and become sharper with decreasing dissipation strength $\alpha$, so that in the limit $\alpha \to 0$, true steps appear in the I-V characteristic. The amplitude of the steps oscillates with the strength of the microwave power as can be seen from Eqs.~(\ref{I-monochromatic}). We find that at large microwave power, the current can become negative at positive voltage demonstrating ANC [Figs.~\ref{FigShapirosteps} (a),(d)]. In particular, for microwave amplitudes around $\beta \approx 2.4$ (corresponding to $J_0(\beta) \approx 0$), the ANC occurs over a wide range of $V_{dc}$ [Fig.~\ref{FigShapirosteps} (d)]. These features provide a clear signature of the ac Josephson effect, 
and their observation would help to unambiguously confirm the existence of an exciton condensate in 
$\nu_T=1$ QH bilayers.  Based on the above calculations, we conclude that Shapiro steps are a more robust experimental signature of the Josephson effects than the Fraunhofer diffraction pattern, and they can be observed also in the presence of strong fluctuations of the vortex field. 

We have also studied the dependence of the Shapiro steps and the ANC on the in-plane magnetic field. The main effect of the in-plane magnetic field is to suppress the Josephson-like tunneling and hence the height of  Shapiro steps [Fig.~\ref{FigShapirosteps} (c)]. In the absence of microwaves, the in-plane magnetic field also gives rise to a resonance at voltage  $eV_{dc}=\hbar u Q_B$ due to the Goldstone mode, so that  one might expect to see resonant features at voltages $eV_{dc}=\hbar u Q_B+n\hbar \omega$. However, these wide resonances appear in a large range of voltages, and even in the absence of microwave radiation, can be clearly observed only by studying the second derivative $d^2 I/dV^2$. On the other hand, we find that in the ANC parameter regime the wide resonances appearing in the static I-V characteristic in the presence of the in-plane field can be used to reverse the direction of the dc current [Fig.~\ref{FigShapirosteps} (d)].

In addition to  Shapiro steps, characteristic resonant features can also appear in the real and imaginary parts of the microwave dynamical conductance (see Fig.~\ref{Dyncond}),   calculated using Eqs.~(\ref{I-monochromatic}) for different values of the dissipation strength. As a function of $V_{dc}$,  the real part of the dynamical conductance shows a dispersive profile around the resonance $eV_{dc}=\hbar \omega$, whereas the imaginary part shows a peak, when this condition is satisfied. Similarly as in the case of the I-V characteristic, for stronger amplitudes of the microwave field replicas of resonant features appear at voltages $eV_{dc}=n \hbar \omega$ ($n=0, 1, 2,...$).

Finally, we compare the Shapiro steps and the resonant microwave response in bilayer exciton condensates to similar effects appearing in other  tunneling systems. First, it is known that similar qualitative features, including resonant features in the I-V characteristic and the dynamical conductance at $eV_{dc}=n\hbar\omega$, can occur in semiconductor heterostructures due to the photon-assisted tunneling \cite{Wacker,Platero,Photon-assisted-transport}. However, important quantitive differences occur because the resonant features considered in this work appear due to the collective dynamics of the exciton condensate. Whereas the effects due to the photon-assisted transport can typically be seen in a THz frequency range in semiconductor heterostructures \cite{Photon-assisted-transport}, we predict that in QH exciton condensates the Shapiro  steps exist already at GHz frequencies. Secondly, we would like to elaborate the analogy to the Shapiro steps in JJs.  For JJs it is possible to show rigorously that the static I-V characteristic can be calculated using lowest order perturbation theory in the limit of large fluctuations, whereas the limit of vanishing fluctuations can be described by means of perturbation theory to all orders in the tunneling \cite{Ingold,IvanchenkoZilberman}. Moreover, as a function of the decreasing temperature, the I-V characteristic develops from the perturbative limit with finite resistance smoothly to the usual Josephson supercurrent I-V characteristic with a critical current given by the Ambekaokar-Baratoff relation \cite{Ingold, IvanchenkoZilberman, Steinbach}. It is known that the microwave response in JJs in the presence of strong fluctuations  is described by the Tucker relations [Eqs.~(\ref{I-monochromatic})]  \cite{Falci}, where the electron charge $e$ is replaced by Cooper pair charge $2e$. In particular, the Tucker relations predict that the heights of the Shapiro steps are proportional to $J_n^2(2eV_\omega/\hbar\omega)$. On the other hand, in ideal JJs  the I-V characteristic is described by Werthamer relations \cite{Werthamer, Hamilton}, where the heights of the Shapiro steps are proportional to $J_n(2eV_\omega/\hbar\omega)$. Based on the analysis of the static I-V characteristic, we expect  a smooth transition between these two regimes as a function of temperature. Our calculations in QH exciton condensates are in the experimentally relevant regime where $\Delta_{SAS}$ is small and the fluctuations of the vortex field are strong, so that the tunneling can be treated using the perturbation theory \cite{Hyart}. Based on the analogy to JJs, we expect that if the temperature is decreased or $\Delta_{SAS}$ is increased, quantitative differences to our results will occur. However, similarly as in the case of JJs we may expect that the results obtained using the perturbation theory are qualitatively correct in all parameter regimes.

In summary, we have studied 
the interlayer tunneling in QH bilayers in the presence of microwave radiation. We have shown that the microwave radiation gives rise to Shapiro-like steps in the tunneling I-V characteristic and we have predicted resonant features in the microwave dynamical conductance. According to our theoretical calculations, the Shapiro steps are more robust experimental signature of the Josephson-like dynamics in bilayer exciton condensates than the Fraunhofer diffraction pattern, and they can be observed also in the presence of strong fluctuations of the vortex field.

We would like to acknowledge valuable  discussions with  W.~Dietsche,  K.~von Klitzing, J.~Smet, X.~Huang and  D.~Zhang.


\end{document}